\documentclass[twocolumn]{jpsj3}
\usepackage{txfonts}
\usepackage{bm}
\usepackage{color}
\usepackage{amsmath,amssymb,physics}
\usepackage{color}

\newcommand{\cket}[1]{|#1\rangle}

\title{Deriving quantum spin model for a zigzag-chain ytterbium magnet with anisotropic exchange interactions}
\author{Hidehiro Saito\thanks{saito-hidehiro722@g.ecc.u-tokyo.ac.jp}, Hiroki Nakai, and Chisa Hotta}
\inst{Department of Basic Science, University of Tokyo, 3-8-1 Komaba, Meguro, Tokyo 153-8902, Japan}

\abst{
We derive a quantum spin Hamiltonian of the spin-1/2 zigzag chain realized in 
a rare earth ytterbium-based magnetic insulator, YbCuS$_2$. 
This material undergoes a transition at 0.95K to an incommensurate magnetic phase with small moments, which does not conform to the nonmagnetic singlet ground state of the spin-1/2 Heisenberg model. 
We take account of octahedral crystal field effect, atomic spin-orbit coupling, and strong Coulomb interactions on Yb ions, 
and perform a four-order perturbation theory to evaluate the superexchange coupling constants. 
A small but finite anisotropic exchange coupling called $\Gamma$-term appears 
similarly to the case reported previously in other triangular-based magnets. 
By varying several material parameters, we figure out two important factors to enhance $\Gamma$-term. 
One is the splitting of excited $f$-states with two holes, which efficiently selects the perturbation terms 
associated with the lowest excited state having large total angular momentum, and accordingly with high spatial anisotropy. 
The other is the tilting of octahedra or distortion of S-Yb-S bond angle, which break the symmetry of the Slater-Koster overlap. 
These effects are systematically analyzed by the exchange anisotropy in units of pairs of octahedra within the 
local spin frame, which significantly reduces the complexity of directly referring to the Hamiltonian discussed in the global axis. 
}
\begin{document}
\maketitle
\section{Introduction}
Intermetallic compounds based on rare earth ions from Ce ($4f^1$) to Yb ($4f^{13}$) show a variety of phases that have long been discussed in the context of valence fluctuation, heavy fermionic and dense Kondo behavior\cite{Stewart2001,Onuki2004,Lohneysen2007,Gegenwart2008}. 
These phenomena are attributed to the particular feature of $4f$ electrons; the outer $5s$ and $5p$ electrons screen them and weaken the effect of the crystal field to be comparable or smaller than the spin-orbit interactions. 
As for the compounds with commensulate electron filling factors, they can easily form an insulator, because of a strong Coulomb interaction compared to that of transition metals. 
For example, in Yb$_4$As$_3$ the electrons become a charge-ordered insulator and its spin degrees of freedom form a one-dimensional antiferromagnet\cite{Kohgi1997}. 
However, unlike the standard quantum magnets based on transition metals, such rare-earth magnetism cannot be explained by the simple Heisenberg model, because the interplay of strong spin-orbit coupling and crystal field generates substantial anisotropic magnetic interactions. 
Indeed, for Yb$_4$As$_3$ the Dzyaloshinski-Moriya (DM) interactions and the staggard spin anisotropic terms are considered to be responsible for the field-induced spin gapped behavior\cite{Oshikawa1999}. 
\par
Compared to other magnets like iridates and ruthenates also with strong spin-orbit coupling, 
ytterbium-based compounds may provide a simpler platform to study anisotropic exchange interactions\cite{Rau2018}. 
This is because the effective spin-1/2 Kramers doublet of Yb$^{3+}$ has 
higher crystal field excitation energies than other rare-earth metal ions\cite{Bertin2012}. 
Previously, toward the realization of quantum spin ice\cite{Ross2011,Gingras2014,Rau2019}, 
the effect of anisotropic exchange interactions in pyrochlore compounds such as Yb$_2$Ti$_2$O$_7$\cite{Hallas2018} 
and in the breathing pyrochlore compound Ba$_3$Yb$_2$Zn$_5$O$_{11}$\cite{Kimura2014,Haku2016,Rau2016} 
has attracted a great deal of attention. 
The case of triangular lattice has also been intensively investigated\cite{Li2016,Zhu2018,Maksimov2019}, 
motivated by the spin liquid-like behavior of YbMgGaO$_4$\cite{Li2015_1,Li2015_2} 
and delafossites $A$Yb$Ch_2$ ($A$ = alkali metal, $Ch$ = chalcogen)\cite{Baenitz2018, Ranjith2019, Schmidt2021} 
with perfect triangular geometry. 
There are several other Yb-based insulating magnets like kagome magnet Yb$_3$Mg$_2$Sb$_3$O$_{14}$\cite{Dun2017}, 
honeycomb compound YbOCl\cite{Zhang2022}, and hyperkagome material Li$_3$Yb$_3$Te$_2$O$_{12}$\cite{Khatua2022}. 
\par
Among them, we focus on YbCuS$_2$ which is recently found to be a good spin-1/2 insulating magnet forming a one-dimensional (1D) zigzag chain. 
The material undergoes a first-order transition at $T_O=0.95$K characterized by the strong divergence of the specific heat\cite{Ohmagari2020}, and at low temperatures, an incommensurate magnetic structure is detected by the neutron scattering experiment. 
In the Cu-NQR study, the $T$-linear behavior of $T_1^{-1}$ is observed at $T<0.5$K\cite{Hori2023}, indicating some sort of gapless excitation. 
In the experimentally obtained magnetic field-temperature phase diagram, the low-temperature phase transforms to an up-up-down (UUD) phase at around 5T. 
The UUD phase typically appears as a magnetization plateau in the spin-1/2 systems on frustrated triangular-based lattices. 
However, theoretically, the spin-1/2 zigzag Heisenberg antiferromagnet hosts a UUD plateau phase when the ground state in a zero-field is a gapped spin singlet phase. 
Therefore, it is naturally expected that YbCuS$_2$ is not a simple Heisenberg antiferromagnet. 
\par
In this work, we derive a quantum spin model for the insulating phase of YbCuS$_2$ by evaluating the exchange interactions between the localized moments of Yb-ions in the presence of strong spin-orbit coupling and crystal field from the octahedral S-ligands. 
The model turns out to deviate from a perfect Heisenberg model. 
In particular, a so-called $\Gamma$-term bond symmetric exchange interaction which was discussed in the Kitaev materials can give leading contributions to the anisotropic exchange of spins. 
%
%
\begin{figure}
\centering
\includegraphics[width=8.5cm]{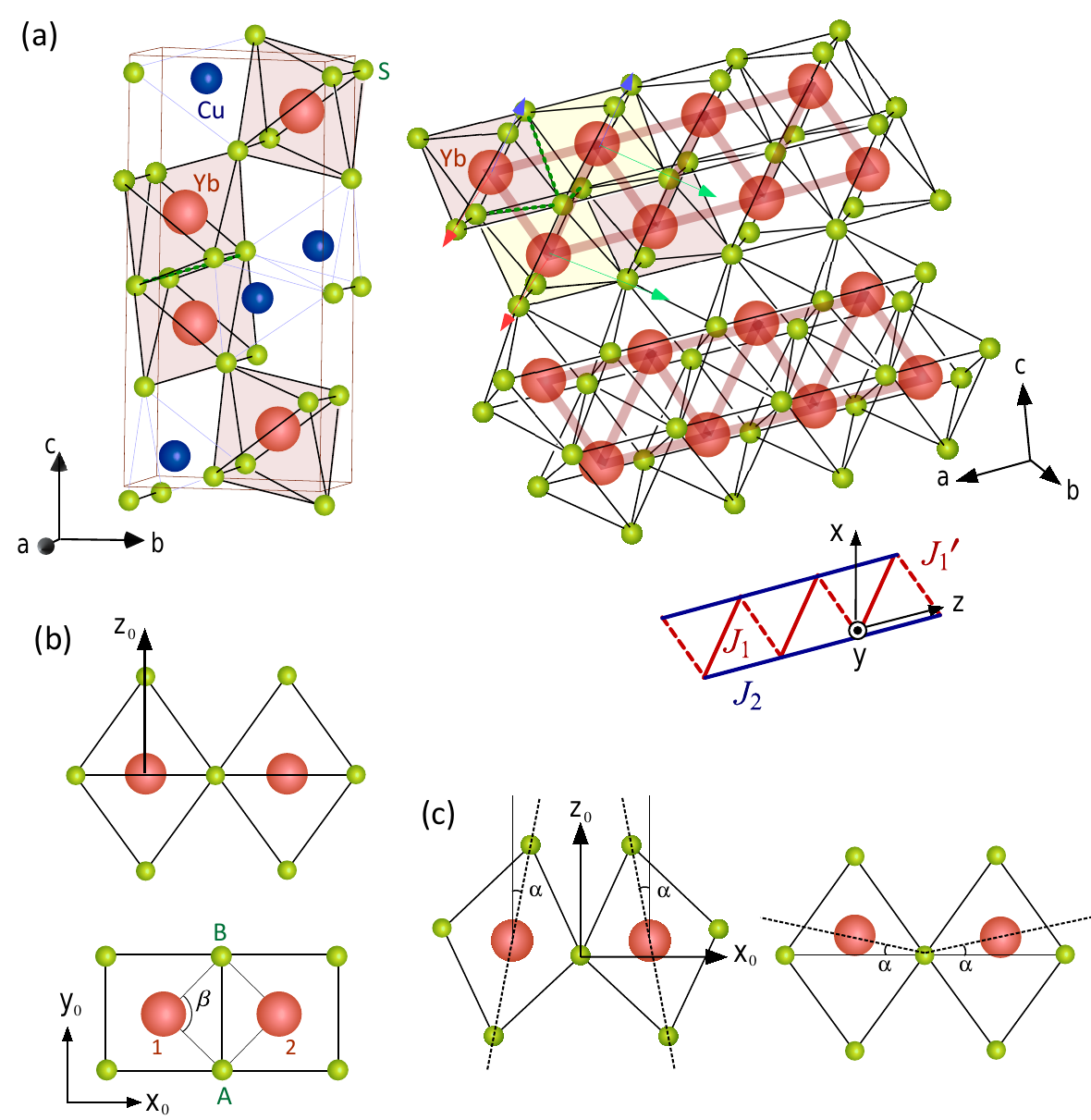}
\caption{(Color online) (a) Crystal structure of YbCuS$_2$. 
Each Yb ion is surrounded by six S ions which gives an octahedral crystal field to the center Yb ion. 
The adjacent octahedrons share an edge, e.g. the shared edges are shown in green broken lines in the left panel. 
The Yb ions form zigzag chains, whose leg runs in the $a$-direction. 
The inter-zigzag chain distances are relatively large and we consider the system to be an ideal 1D zigzag ladder with $J_1$ and $J_2$ bonds. 
The $J_1$ bonds running in two different directions (solid and broken red lines) are crystallographically equivalent. 
(b) Two adjacent octahedra are shown in two different angles. 
We consider the $4f$-orbitals of Yb-1 and Yb-2 and $p$-orbitals of S-A and S-B which have overlap integrals $t_{i\lambda}$ between $i=1,2$ Yb ion and $\lambda=A,B$ S ion. We define the bond angle S-Yb-S as $\beta$.
(c) The case where the crystal has distortion(left panel) and the octahedron is 
off the regular shape(right panel) by angle $\alpha$. 
We find the latter with $\alpha=3.5^\circ$ in YbCuS$_2$\cite{kaneshima2021}. 
}
\label{f1}
\end{figure}
\begin{figure*}[t]
\includegraphics[width=18cm]{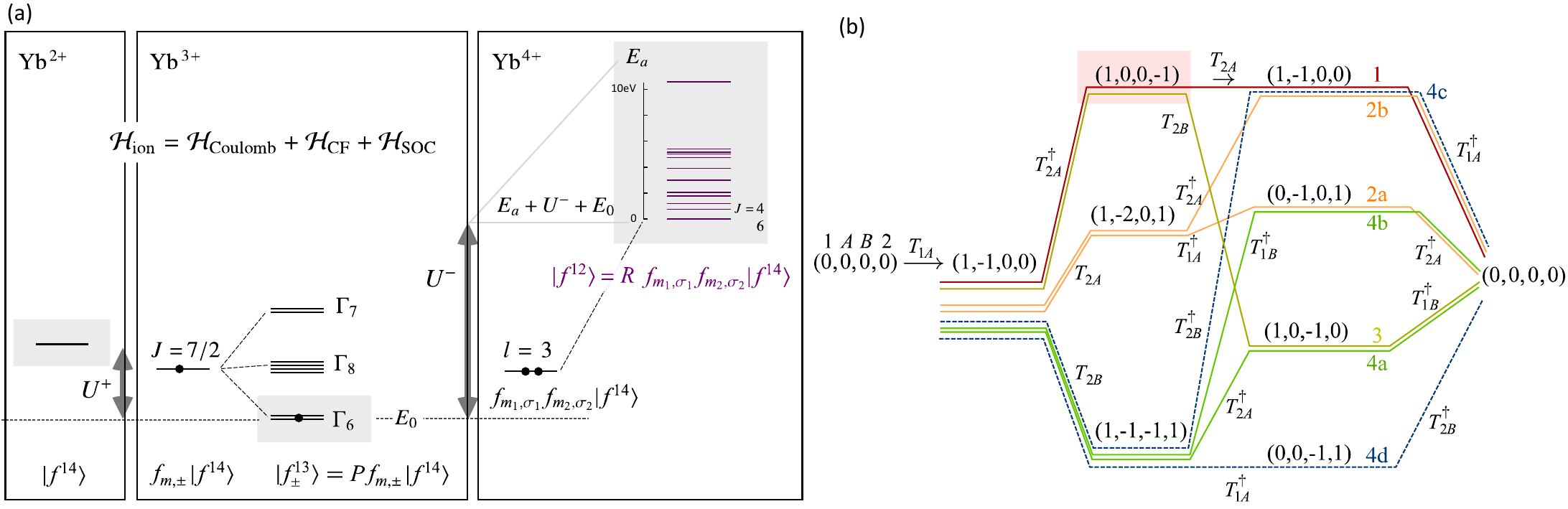}
\caption{(Color online) (a) Energy levels of single Yb$^{2+}$, Yb$^{3+}$, Yb$^{4+}$ ions surrounded by the regular octahedron, obtained as an eigenstate of Eq.(\ref{eq:hion}). 
The highlighted energy levels, $U^++E_0$, $E_0$, $E_a+U^-+E_0$ for each valence, are used in the present calculation. 
(b) Eight independent perturbation processes, denoted as 1, 2a, 2b, 3, 4a, 4b, 4c, 4d, starting from the common process of $(0,0,0,0)\rightarrow (1,-1,0,0)$, where we use the representation of states by electron number counted from $f^{13}$ and $p^6$ ground states of (Yb-1, S-A, S-B, Yb-2) ions. $T_{\gamma\lambda}$ are the $7\times 3$ hopping matrix between $\gamma=1,2$, $\lambda=A,B$ ions obtained based on Slater-Koster tables for $f$-electrons\cite{Takegahara1980}. 
The excited states including the $f^{12}$ levels are highlighted. 
}
\label{f2}
\end{figure*}
\section{Crystal structure and local electronic states}
\subsection{Crystal structure}
The crystal structure of YbCuS$_2$ is shown schematically in Fig.~\ref{f1}(a). 
Each Yb ion is surrounded by the S$_6$ octahedron, and the octahedrons share their edges in the $a$-direction, while in the $c$-direction, every two octahedrons share edges. 
As a result, the Yb ions mediated by the sulfur ions form a zigzag chain. 
In deriving the effective Hamiltonian, the global coordinate has the $z$ axis along the leg of the chain and the triangular plane lines on the $ zx$ plane, as shown in the figure. 
\par
In evaluating the superexchange interactions $J_1$ and $J_2$ between the magnetic moments of the Kramers doublet on adjacent Yb-ions, we prepare a local $x_0y_0z_0$-axis to represent the crystal field on each octahedron as shown in Fig.~\ref{f1}(b) 
for each pair of octahedra, independently for different bonds. 
When the octahedra are slightly distorted, the S-Yb-S angle denoted as $\beta$ deviates from 90$^\circ$, 
and the two octahedra may tilt their orientation by angle $\alpha$ as shown in Fig.~\ref{f1}(c). 
In YbCuS$_2$, we find structures where Yb is slighly off the center of the octahedron (right panel of Fig.~\ref{f1}(c)) by  
$\alpha \sim 3.5^\circ$. 
\par
To describe the $f$-electron states on these atoms, we prepare a set of bases with angular momentum $l=3$ using this 
local coordinate as $\big\{\;\cket{x_0y_0z_0},\,\cket{x_0(5x_0^2-3r^2)},\,\cket{y_0(5y_0^2-3r^2)},\,\cket{z_0(5x_0^2-3r^2)},
\,\cket{x_0(y_0^2-z_0^2)},\,\cket{y_0(z_0^2-x_0^2)},\,\cket{z_0(x_0^2-y_0^2)} \;\big\}$ and combine it with the 
spin momentum to describe the multiplet states in the presence of spin-orbit coupling 
(see Appendix B). 

\subsection{Single ion energy levels}\label{sec:f14}
We take a ``vacuum" of the $f$-electron state of a single Yb ion as $|f^{14}\rangle$ representing a closed shell of $4f$-orbital realized for divalent Yb$^{2+}$ state. 
To deal different valence states on equal footing, we introduce the creation operator $f_{m,\pm}$ of $f$-hole (annihilation operator of $f$-electron) with angular momentum $l=3$ indexed as $m=-3,\cdots,3$ and spin-1/2 momentum $s^z=\pm 1/2$, because the electron states with other angular momentum do not participate in the low energy energetics. 
The $f^{14}$ state is given by adding these electrons to $|0\rangle$ to be fully occupied in the order as 
\begin{equation}
|f^{14}\rangle=\prod_{\sigma=\pm}\big(\prod_{m=-3}^{3} f^\dagger_{m,\sigma}\big) |0\rangle. 
\end{equation}
\par
In crystals, ytterbium tends to form trivalent and divalent states and the valence fluctuation between them in many intermetallic compounds. 
Therefore, it is natural to expect that the energy levels of an isolated Yb$^{2+}$ and Yb$^{3+}$ ions are close to each other. 
\par
We first explain the overall energy diagrams given in Fig.~\ref{f2}(a). 
The Hamiltonian for the isolated Yb-ions with SOC and in a crystal field formed by the S-octahedron consists of three terms, 
\begin{equation} 
{\mathcal H}_{\rm ion}={\mathcal H}_{\rm Coulomb}+ {\mathcal H}_{\rm CF} + {\mathcal H}_{\rm SOC}, 
\label{eq:hion}
\end{equation}
and their energy scales are 10eV, 0.1eV, and 0.1eV, respectively.
Their eigenstates largely depend on the valence of the ion, particularly the term, ${\mathcal H}_{\rm Coulomb}$. 
For Yb$^{2+}$, closed shell of $|f^{14}\rangle$, the last two terms do not matter because they only shift the energy levels. 
For Yb$^{3+}$, a set of $\{ f_{m,\pm}|f^{14}\rangle\} $, mix by ${\mathcal H}_{\rm CF} + {\mathcal H}_{\rm SOC}$ and form one-body eigenstates of ${\mathcal H}_{\rm ion}$, whose lowest state with energy $E_0$ ($\Gamma_6$ in Fig.~\ref{f2}(a)) is the ground state. 
The energy difference is denoted as $U^+=\langle f^{14}| {\mathcal H}_{\rm ion} |f^{14}\rangle - E_0$. 
\par
For Yb$^{4+}$, we have two holes, and solving ${\mathcal H}_{\rm ion}$ including the two-body term, ${\cal H}_{\rm Coulomb}$, will yield $\,_{14}C_2=91$ eigen states, whose energy is denoted as $E_a+U^-+E_0$, where $U^-$ is the lowest energy measured from $E_0$. 
Because Yb$^{4+}$ is much unstable than Yb$^{2+}$ and Yb$^{3+}$, $U^-\gg U^+ >0$ is expected. 
\subsection{Single ion: $f^{13}$ state}\label{sec:f13}
Let us consider an isolated unit of Yb$^{13}$ atom and its surrounding octahedron, which serves as a starting point of the perturbation theory. 
Each Yb$^{3+}$ ion has $4f^{13}$ electron configuration which corresponds to having one hole, and due to strong SOC, they form a low-lying $J=l+s=7/2$ multiplet consisting of angular momentum $l=3$ and spin momentum $S=1/2$. 
The octahedral crystal field further splits this 8-fold multiplet into $\Gamma_7$, $\Gamma_8$ and $\Gamma_6$ levels, and the $4f^{13}$-ground state is in the lowest $\Gamma_6$ Kramers doublet. 
Taking the quantization-axis of the octahedron as $z_0$-axis shown in Fig.~\ref{f1}(b), this doublet is given by 
\begin{equation}
|f^{13}_{\pm}\rangle =\sqrt{\frac{5}{12}}|\pm 7/2\rangle + \sqrt{\frac{7}{12}}|\pm 1/2\rangle 
\label{eq:kramers}
\end{equation}
where we denote the states on the R.H.S. by the $z$-element of the coupled angular momentum as $|J^z\rangle$. 
When spanned by a set of single-hole states, they are expressed as 
\begin{align}
|f^{13}_+\rangle &= \Big(-\frac{1}{\sqrt{3}} f_{0,+}- \sqrt{\frac{5}{12}} f_{-3,-} -\frac{1}{2}f_{1,-} \Big)\:|f^{14}\rangle 
\nonumber\\
|f^{13}_-\rangle &= 
\Big(\frac{1}{2} f_{-1,+}+ \sqrt{\frac{5}{12}} f_{3,+} + \frac{1}{\sqrt{3}}f_{0,-} \Big)\:|f^{14}\rangle ,
\label{eq:single-f13}
\end{align}
which are expressed as $|f^{13}_\pm\rangle = \sum_{\mu} \big(\hat P\big)_{\pm,\mu} \big(f_{\mu}|f^{14}\rangle\big)$ with $\mu=(m,\sigma)$, using the $2\times 14$ matrix $\hat P$. 
\subsection{Single ion: $f^{12}$ states}\label{sec:f12}
For excitation from the $f^{13}$-Kramers doublet, we need to consider $f^{12}$-state consisting of two holes. 
They are spanned by a set of 91 states indexed by the angular and spin momentum of holes as, $|m_1\sigma_1;m_2\sigma_2\rangle = f_{m_1,\sigma_1}f_{m_2,\sigma_2}|f^{14}\rangle$, where $(m_1,\sigma_1)\ne (m_2,\sigma_2)$. 
We obtain an eigen state of Eq.(\ref{eq:hion}). 
Because we have two holes, the Coulomb interaction between them works 
not as a constant shift of the energy levels but to split them, 
which is expressed using the operator $f_{m\sigma}$ as 
\begin{align}
{\mathcal H}_{\rm Coulomb}&= \sum_{k=0,2,4,6} a_k F^k \sum_{q=-k}^k O_{kq}^\dagger O_{kq}, \\
& O_{kq}= \sqrt{\frac{2l+1}{2k+1}} \sum_\sigma\sum_{mm'}(-)^m  \langle l,-m,l,m'|k,q\rangle
  f_{m\sigma}^\dagger f_{m'\sigma}, 
\label{eq:hcoulomb}
\end{align}
with $\langle l,-m,l,m'|k,q\rangle$ being the Clebsch-Gordan coefficient with $l=3$ and 
\begin{equation}
{\mathcal H}_{\rm SOC}= \zeta \sum_{\sigma\sigma'}\sum_{mm'} \big(\bm l_{mm'}\cdot \bm s_{\sigma\sigma'}\big)
f_{m\sigma}^\dagger f_{m'\sigma'}, 
\label{eq:hsoc}
\end{equation}
with $\bm l$ and $\bm s$ being the matrix representation of $l=3$ angular momentum and spin $s=1/2$ momentum, respectively. 
In evaluating these matrices, we use $(a_2,a_4,a_6)=(2/15,1/11,50/429)$ and $F^2=14.184$eV, $F^4 = 9.846$ eV and $F^6 = 6.890$ eV for the Coulomb interactions\cite{meftah2013}, and 
$\zeta= 0.380$eV for the SOC. 
Here, we discard the term, $F^0$, which gives the constant shift to the energy level, which is included in the $U^-$ term. 
Accordingly, the lowest energy state is set to zero for $E_a$. 
\par
Diagonalizing ${\mathcal H}_{\rm ion}$ we obtain 13 different eigen energies $E_a$ and corresponding 91 eigenstates. 
The energy levels are listed in Table~\ref{tab1} in Appendix~\ref{sec:f12_2}. 
The corresponding 91 eigenstates are given as 
$|f^{12}_a\rangle = \sum_\mu \big(\hat Q\big)_{a,\mu} \big(f_{\mu}|f^{14}\rangle\big)$ with $\mu=(m,\sigma)$, 
using the $91\times 14$ matrix $\hat Q$. 
\subsection{Slater-Koster parameters}\label{sec:slater}
Let us take the common local $x_0y_0z_0$-axis for two adjacent two octahedra which share the edges as shown in Fig.~\ref{f1}(c) 
(we set $\alpha=0$). 
The direct overlap of the two Yb-orbitals is small and neglected, and we consider the overlaps between $p$-orbitals $\big\{\cket{p_x}, \cket{p_y}, \cket{p_z}\big\}$ of S-ions on both sides of the shared bonds and of the isolated $f$-orbitals labeled by angular momentum, $m=-3,\ldots,3$. 
The overlap integral matrices are evaluated using the Slater-Koster table\cite{Takegahara1980} and taking the direction cosines 
$(\ell,m,n)=(d_{12}/\sqrt{d_{12}^2+d_{AB}^2},\mp d_{AB}/\sqrt{d_{12}^2+d_{AB}^2},0)$ with 
$d_{12}$ and $d_{AB}$ being the distance between Yb-1 and Yb-2 ions, and sulfer A and B ions, respectively. 
\par
The $7\times 3$ hopping matrix between Yb ($\gamma=1,2$) and S($\lambda=A,B$) ions is given as $\hat T_{\gamma\lambda}= V_{pf\sigma}\hat T_\sigma+ V_{pf\pi} \hat T_\pi$, which includes two terms, $\hat T_\sigma$ and $\hat T_\pi$, representing the $\sigma$ and $\pi$-overlaps, respectively. 
The ratio of constant overlap parameters is typically set as $V_{pf\pi}/V_{pf\sigma}=-0.3$. 
For example, $(\hat T_\sigma)_{p_x,x_0y_0z_0}=\sqrt{15}l^2 mn$ and $(\hat T_\pi)_{p_x,x_0y_0z_0}=-\sqrt{5/2}(3l^2-1) mn$. 
When we have a tilting of angle $\alpha \ne 0$, the overlaps of these functions is evaluated using $(\ell,m,n)$ different from above. 
\par
For cases where the crystals deviate from the combination of perfect octahedra, 
we consider two angles $\alpha,\beta$ shown in Figs.~\ref{f1}(b),(c). 
For simplicity we do not consider the modification of crystal field but focus on their effect on the Slater-Koster 
parameters through the modification of direction cosines, which as shown in detail in Appendix.~\ref{sec:angles}. 
%
%
\section{Perturbation theory}
\subsection{Fourh order processes}
We now evaluate the superexchange interactions between magnetic moments on two adjacent Yb-ions shown in Fig.~\ref{f1}(b) labeled as $\gamma=1,2$. 
We follow most of the notations of Ref.~\cite{Rau2018} that provide the same types of perturbations on Yb-based materials, while we add several terms that are discarded in their treatment and examine the parameter dependences for the zig-zag chain. 
\par
The unperturbed Hamiltonian is given as ${\mathcal H}_0 \equiv {\mathcal H}_{f,1}+{\mathcal H}_{f,2}+{\mathcal H}_{p,A}+{\mathcal H}_{p,B}$, where the first two terms are for the $f$-electrons on isolated Yb-ions, and the latter two are those for $p$-electrons on two S-ions labeled as $\lambda=A,B$. 
\par
The perturbation is given by the hoppings between $f$ and $p$-orbitals given as 
\begin{equation}
{\mathcal H}' 
= \sum_{\mu \nu} \sum_{\gamma=1,2} \sum_{\lambda=A,B} [ \big(\hat T_{\gamma\lambda}\big)_{\mu \nu}
f^{(\gamma)\dagger}_{\mu} p^{(\lambda)}_{\nu} + \mathrm{H.c.}], 
\label{eq:hprime}
\end{equation}
where $f^{(\gamma)}_{\mu}$ with index $\mu=(m,\sigma)$ representing the $f$-electron with the orbital $m=-3,\cdots,3$ and spin momentum $\sigma=\pm$, and $p^{(\lambda)}_{\nu}$ is the anihilation operator whose index $\nu$ represents the electron on $p$-orbital ($\eta= x_0,y_0,z_0$) with spin $\sigma$ of S-ion on site $\lambda$=A,B. 
The transfer integrals $\big(\hat T_{\gamma\lambda}\big)_{\mu \nu}$ are the ones given in \S. \ref{sec:slater}. 
\par
The fourth order perturbation is expressed using the operator $P$ and $R$ that represents the transformation from the $l=3$ basis to the ground multiplet of $f^{13}$ and the eigenstates of $f^{14}$ or $f^{12}$ states given earlier, respectively. 
\begin{align}
{\cal H}_{\rm eff}&= P^\dagger {\cal H}' Q {\cal H}'  Q  {\cal H}' Q {\cal H}'  P, \nonumber \\
& \tilde Q = R^\dagger (\Delta {\cal H}_{\rm ions})^{-1} R
\end{align}
where $\Delta {\cal H}_{\rm ions}$ is the sum of four single ions Hamiltonian measured from $2(E_0+E_p)$ where $E_0$ and $E_p$ are the energies of $f^{13}$ and $p^6$, respectively. 
For example, during process 3, we encounter three excited states with $\langle \Delta {\cal H}_{\rm ions} \rangle= U^++\Delta$ for $(1,-1,0,0), (1,0,-1,0)$ states and $U^++U^-+E_a$ for $(1,0,0,-1)$, where $\Delta$ is the energy per hole on the S-ion, where we label the states as by their electron numbers on $(1,A,B,2)$ measured from the ground state, $(0,0,0,0)$. 
\par
Figure~\ref{f2}(b) shows eight perturbation processes classified by the electron numbers of the excited states, that start by transferring one electron from A to Yb-1, where we have $(1,-1,0,0)$ after operating ${\mathcal H}'$ once. 
Among them processes 4c and 4d perfectly cancel out \cite{Rau2018}, 
which we do not need to take account of. 
Because the system is symmetric about the exchange between A and B as well as 1 and 2, 
there are 24 processes obtained by the symmetry operation.  
\par
We briefly summarize the types of processes; 
In processes 1 and 2, only one S-ion (A or B) takes part. 
Processes 3 and 4 are ring exchange processes where the electrons circulate the 1-A-2-B loop. 
Processes 1 and 3 include $f^{12}$ excited state but processes 2 and 4 do not. 
%
\begin{figure}
\centering
\includegraphics[width=8cm]{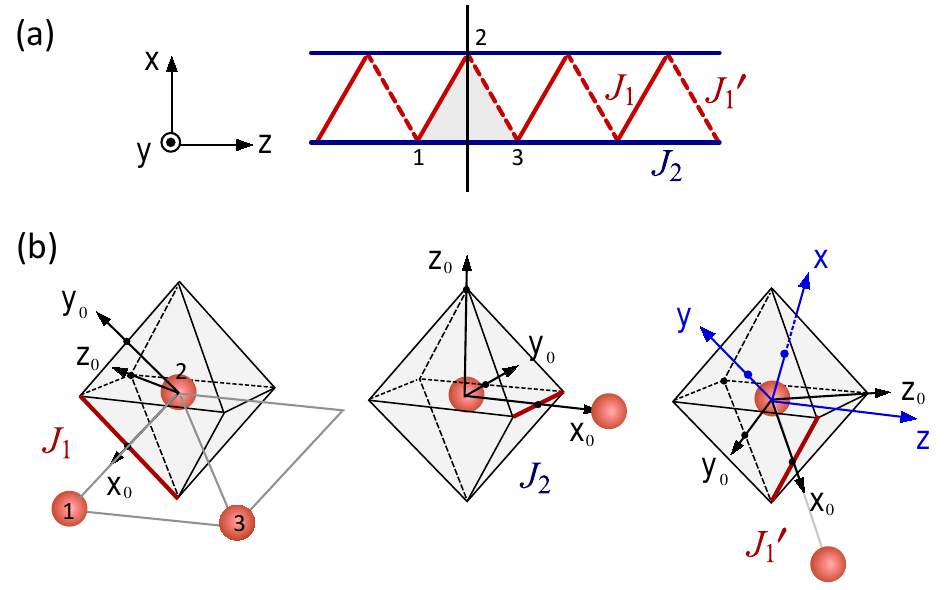}
\caption{(Color online) (a) Schematic illustration of the zigzag chain. 
(b) Local $x_0y_0z_0$-coordinates defined on an octahedron in calculating the exchange interactions in the $J_1, J_1'$, and $J_2$ directions.
}
\label{f3}
\end{figure}
\subsection{Effective spin Hamiltonian}
By summing up all the fourth-order perturbation processes, 
the two-spin Hamiltonian on $J_b \:(b=1,1',2)$ bonds (see Fig.~\ref{f3}) 
are obtained whose spin representation is defined independently as $x_0y_0z_0$-axes (which can even differ between sites $\gamma=1,2$ on the same bond when the octahedra are tilted) as, 
\begin{equation}
h^{(b)}_{\rm eff}= 
\sum_{\mu,\,\nu=x,y,z}\; S^\mu_{0;1} \;\big({\hat J}_{0;b}\big)_{\mu\nu}  S^\nu_{0;2}, 
\label{eq:hlocal}
\end{equation}
where $S^\mu_{0;\gamma}$ ($\gamma=1,2$) is the spin-1/2 operator representing the $\Gamma_6$ doublet of Yb-$\gamma$ on the $l$-th bond and ${\hat J}_{0;b}$ is the $3\times 3$ matrix representing the superexchange interaction. 
\par
The symmetry of ${\hat J}_{0;b}$ stores information on what types of processes or parameters influence the 
anisotropy of interactions. 
When we have perfect octahedra, the system has an inversion symmetry and ${\hat J}_{0;b}$ is diagonal, 
while if they tilt or distort as in Fig.~\ref{f1}(c), the inversion symmetry is lost 
and the off-diagonal element, $\pm D$, emerges, as 
\begin{align}
{\hat J}_{0;b}=\left( \begin{array}{ccc} 
 X & 0 & D\\
 0 & Y & 0 \\
-D & 0 & Z \\
\end{array}\right). \;
\label{mat:j0}
\end{align}
\par
Because, $x_0y_0z_0$-axis on the same site belonging to different bond-$b$ are not related, $h^{(b)}_{\rm eff}$ needs to be 
transformed to the global $xyz$-axis using the unitary matrix $\hat U_{\gamma;b}$. 
The two representation of spin operators on Yb-$\gamma$ are related to each other as $S^\mu_{0;\gamma}= \sum_{\nu} \big(\hat U_{\gamma;b}\big)_{\mu\nu} S^\nu_{\gamma}$, and the effective Hamiltonian is given as 
\begin{equation}
{\cal H}_{\rm eff}= \sum_{\langle i,j\rangle\in b}\sum_{\mu,\,\nu=x,y,z}\;
S^\mu_i \:\big({\hat J}_{b}\big)_{\mu\nu}\: S^\nu_j, 
\label{eq:hspin}
\end{equation}
where $\langle i,j \rangle$ is the nearest neighbor pairs of sites and $S^\mu_i$ is the $\mu=x,y,z$ element of spin-1/2 operator defined for the global $xyz$-axis. 
The $3\times 3$ exchange coupling of bond directions, ${\hat J}_b$, are obtained as 
${\hat J}_b =U_{i;b}^{\:T}\: {\hat J}_{0;b}\: U_{j;b}$. 
The transformation yields, 
\begin{align}
{\hat J}_2=\left( \begin{array}{rrr} 
(Y+Z)/2 & (-Y+Z)/2 & -D/\sqrt{2} \\
(-Y+Z)/2 &(Y+Z)/2  & -D/\sqrt{2} \\
D/\sqrt{2} \;\;& D/\sqrt{2}\;\; & X \;\;\\
\end{array}\right). 
\label{mat:j2}
\end{align}
and the symmetric part of ${\hat J}_{1}$ and ${\hat J}_{1'}$ are 
\begin{align}
{\hat J}_{1/1'}^{\rm (sym)}=\left( \begin{array}{ccc} 
\frac{Z}{4}+c_1^2X+c_2^2Y & \\
\frac{X+Y-2Z}{8} & \frac{Z}{4}+c_2^2X+c_1^2Y \\
\mp \frac{Z}{2\sqrt{2}}\pm \frac{c_1 X}{2}\mp \frac{c_2 Y}{2} & \pm \frac{Z}{2\sqrt{2}}\pm \frac{c_2X}{2} \mp \frac{c_1 Y}{2} 
& \frac{X+Y+2Z}{4} 
\end{array}\right), 
\label{mat:j1}
\end{align}
with $c_1=(2+\!\sqrt{2})/4$, $c_2=(2-\!\sqrt{2})/4$, and the antisymmetric part are 
\begin{align}
{\hat J}_{1/1'}^{\rm (asym)}=\left( \begin{array}{ccc} 
 0 & D/2&   \pm c_1 D \\
 -D/2 & 0&   \mp c_2 D \\
 \mp c_1 D & \pm c_2 D & 0 
\end{array}\right). 
\label{mat:j1D}
\end{align}
When Eq.(\ref{mat:j0}) form a Heisenberg interaction with $X=Y=Z$ and $D=0$, 
${\hat J}_{2}$ and ${\hat J}_{1/1'}$ become the same 
and the global Hamiltonian is a Heisenberg model. 
When Eq.(\ref{mat:j0}) form an XXZ interaction, $X=Y \ne Z$ and $D=0$, 
symmetric off-diagonal element called $\Gamma$-term appears in the global Hamiltonian. 
When Eq.(\ref{mat:j0}) form an XYZ interaction $X \ne Y \ne Z$ or 
with a finite Dyaloshinskii-Moriya interaction, $D\ne 0$, 
the model becomes more highly anisotropic. 
\par
Because we mostly deal with the case of $X=Y$, or even if $X\ne Y$ their difference is small, 
we can safely parameterize the degree of symmetric anisotropy as 
\begin{equation}
\Gamma=(-(X+Y)+2Z)/4, 
\label{eq:gamma}
\end{equation}
where $X,Y,Z$, and $\Gamma$ take different values depending on bonds, $b=1,1',2$, 
while we omit the subscript $b$ to avoid confusion with the label of multiplets. 
When $X=Y$, Eq.(\ref{eq:gamma}) is indeed equivalent to $(-Y+Z)/2$, 
and the value of $\Gamma$ on each bond is reflected in the matrix elements of 
the exchange interactions of a global $xyz$-coordinate as 
$(\hat J_2)_{xy}=\Gamma$,
$(\hat J_{1}^{\rm (sym)})_{xy}=\Gamma/2$, and 
$(\hat J_{1}^{\rm (sym)})_{zx}=-(\hat J_{1}^{\rm (sym)})_{zy}=-\Gamma/\sqrt{2}$. 
\par
To view how the global symmetry of the system is reflected in $\hat J_b$, 
let us consider a perfect zigzag chain consisting of edge-shared regular triangles, 
and focus on spins on sites 1,2, and 3 shown in Fig.~\ref{f3}(a). 
The mirror symmetry about the plane perpendicular to $z$-axis is represented by the matrix $\hat M={\rm diag}(1,1,-1)$, and operating it to the triangle we find, 
\begin{equation}
\bm S_1 \rightarrow -M \bm S_3,\;\; 
\bm S_2 \rightarrow -M \bm S_2,\;\; 
\bm S_3\rightarrow -M \bm S_1. 
\end{equation}
Because the Hamiltonian remains invariant by this operation, the exchange interactions need to fulfill 
${\hat J}_1'\,^T= {\hat M} {\hat J}_1{\hat M}$ and 
${\hat J}_2\,^T= {\hat M} {\hat J}_2{\hat M}$. 
Resultantly, $\hat J_1$ and $\hat J_{1'}$ are related to each other as 
\begin{align}
{\hat J}_1=\left( \begin{array}{ccc} 
J_1^{xx} & J_1^{xy} & J_1^{xz} \\
J_1^{yx} & J_1^{yy} & J_1^{yz} \\
J_1^{zx} & J_1^{zy} & J_1^{zz} \\
\end{array}\right), \;
{\hat J}_1'=\left( \begin{array}{crr} 
J_1^{xx} & J_1^{xy} &\!\!-J_1^{xz} \\
J_1^{yx} & J_1^{yy} &\!\!-J_1^{yz} \\
-J_1^{zx} & \!\! -J_1^{zy} &J_1^{zz} \\
\end{array}\right), 
\label{mat:j1shape}
\end{align}
and the elements of $\hat J_2$ needs to take the form, 
\begin{align}
{\hat J}_2=\left( \begin{array}{ccc} 
J_2^{xx} & J_2^{xy} & 0 \\
J_2^{xy} & J_2^{yy} & 0 \\
0 & 0 & J_2^{zz} \\
\end{array}\right). 
\label{mat:j2shape}
\end{align}
Both Eqs.(\ref{mat:j1shape}) and (\ref{mat:j2shape}) are fulfilled 
in Eqs.(\ref{mat:j1}), (\ref{mat:j1D}), and (\ref{mat:j2}) 
regardless of the values of $X,Y,Z,D$. 
%
\begin{figure*}[t]
\centering
\includegraphics[width=17cm]{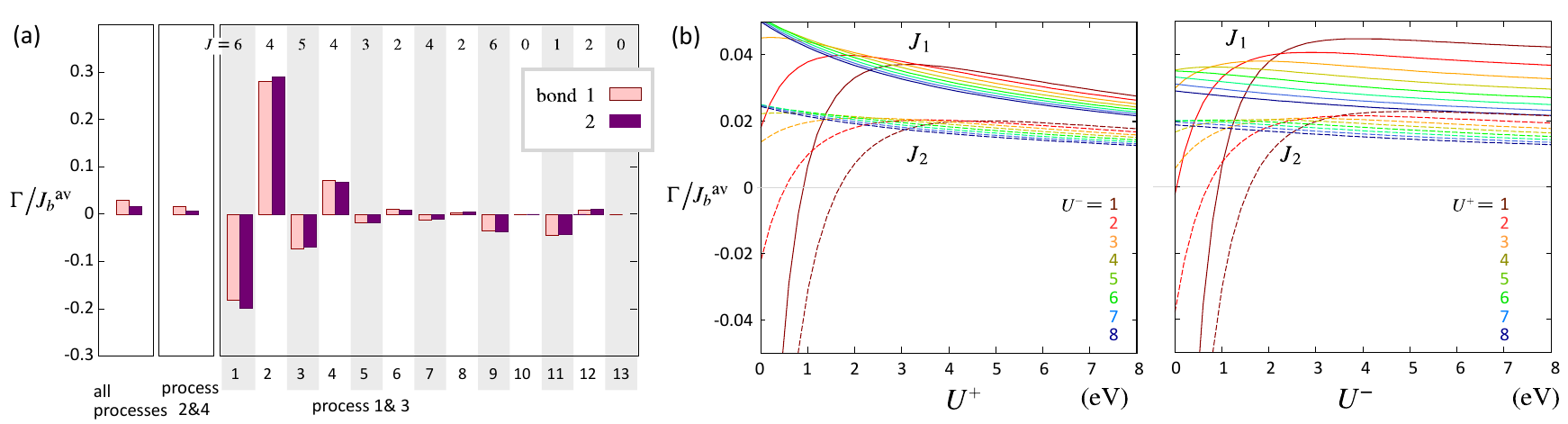}
\caption{(Color online)Results of $\Gamma$ for bonds $b=1,2$ normalized by $J_b^{\rm av}$. 
 (a) Contributions to $\Gamma$ from different processes, 
obtained by setting $U^\pm=5$eV and $\Delta=4$eV (data of Table~I(a)). 
The contributions from process 2 and 4 is shown in the second column, 
and those from processes 1 and 3 passing through 13 energy levels of $E_a$ are shown separately. 
(c) Variation of $\Gamma$ when $U^\pm$ are varied with $\Delta=4$eV. 
}
\label{f4}
\end{figure*}
%
\begin{figure}[tbp]
\centering
\includegraphics[width=8.5cm]{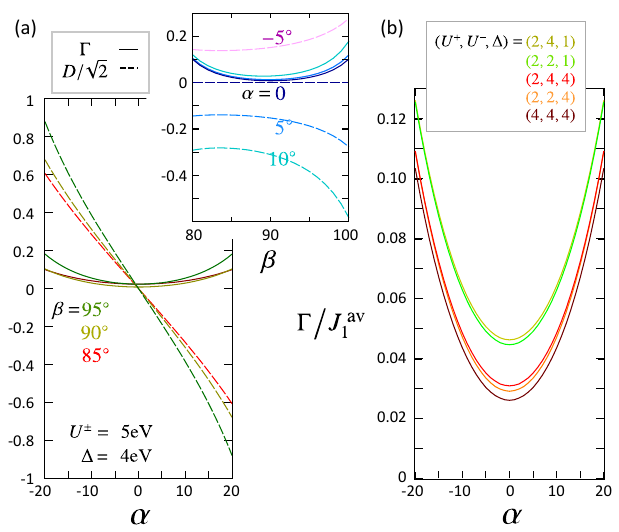}
\caption{(Color online) Results of $\hat J_1$ for the tilted($\alpha\ne 0$) and distorted ($\beta\ne 90^\circ$) 
octahedra (see Fig. 1(b,c)). 
(a) $\Gamma$ and $D/\sqrt{2}$ as a function of $\alpha$ with fixed $\beta=85^\circ,90^\circ,95^\circ$. 
Inset shows $\beta$-dependence for several fixed values of $\alpha$. 
(b) $\Gamma/J_1^{\rm av}$ of bond 1 as a function of $\alpha$ with fixed $\beta=85^\circ$  
for different choices of parameteters, 
$(U^+,U^-,\Delta)=(2,2,1),(2,4,1),(2,2,4),(2,4,4),(4,4,4)$. 
}
\label{f5}
\end{figure}
%
\begin{table}[tbp] \label{tab:jvalues}
\caption{(a) Exchange couplings, $\hat J_{0;b}$, on bonds $b=1,2$ of a local $x_0y_0z_0$ axis, 
and $\hat J_{b}$, transformed to global $xyz$-axis, 
when we set $U^\pm=5$eV, $\Delta=$4eV. 
(b) Symmetries of the exchange coupling, XXX/XXZ/XYZ, for different conditions. 
When we set $E_a=0$ (charging approximation), and the S-Yb-S bond angle $\beta= 90^\circ$, 
and tilting angle $\alpha=0$, we find a Heisenberg model (XXX). 
The breaking of these symmetries brings anisotropy to the exchange coupling. 
For processes 3 $\&$ 4, those going through the levels with total angular momentum $J\ne 0$ (see Table~\ref{tab1}) 
lowers the symmetry while the levels with $J=0$ do not lower the symmetry (XXX) because they have symmetries about the up/down spins. 
}
\begin{tabular}{ccccccc}
\hline
(a)  \rule{2mm}{0mm}&  $xx$ & $yy$ & $zz$ &  $xy$ & $yz$ & $zx$  \\
\hline
$\hat J_{0;1}$ &  0.9816 & 0.9743 & 0.9447 &
\rule{0mm}{3mm}\\
$\hat J_{0;2}$ &  0.8626 & 0.8518 & 0.8080 \\
\hline
$\hat J_1$  \rule{0mm}{3mm}
&  1.0279 & 1.0279 & 1.0209 & -0.0098 &  -0.0098 &  0.0098 \\
$\hat J_2$  &  0.9595 & 0.9595 & 0.9816 & -0.0149 &   0.0000 &  0.0000 \\   
\hline
\end{tabular}
\begin{tabular}{ccccc}
\hline
(b)\:ch. approx. \!\!& \!\!bond symm. & $J_{0;b}$  & pr. 1 $\&$ 3 & pr. 3 $\&$ 4
\\
\hline
\; $E_a=0$ & $\alpha=0,\;\:\beta=90^\circ$ & XXX & XXX & ---  \\   
\; $E_a=0$ &  $\alpha\ne 0\:/\:\beta=90^\circ$ & XXZ & XXZ & ---  \\
\; $E_a\ne 0$ & $\alpha=0,\;\beta=90^\circ$ & XXZ & XXX & XXZ($J\ne 0$) \\   
\; $E_a\ne 0$ & $\alpha\ne 0 \:/\: \beta=90^\circ$ & XYZ & XXZ & XYZ($J\ne 0$) \\   
\hline 
\end{tabular}
\end{table}
\section{Results}
The degree of symmetric and antisymmetric types of anisotropies of exchange interactions are measured by 
$\Gamma$ in Eq.(\ref{eq:gamma}) and $D$, respectively.  
We consider two cases, $\alpha=0$ and $\alpha\ne 0$ which yields $D=0$ and $D\ne 0$, respectively, 
and examine the condition to enhance the anisotropy of exchange interactions. 
As mentioned, we take $V_{pf\pi}/V_{pf\sigma}=-0.3$ and 
set the unit of $\hat J_{b}$ as 10$^{-4}$ eV when setting $V_{pf\sigma} = 1$ eV, 
which is of the order of transition temperature $T_O\sim 1K$. 
However, because of the uncertainty of $V_{pf\sigma}$ and $U^\pm, \Delta, U_p$, 
the amplitudes of $\hat J_{b}$ do not have any physical implication. 
Therefore, we scale the parameters by the average of diagonal exchange interactions as $J_b^{\rm av}=(X+Y+Z)/3$. 
\par
In the following, the material parameters to be varied are $U^\pm, \Delta, U_p$ and $E_a$, 
while among them only process 2 has contributions from $U_p$ and the influence is small 
so that we fix it to $U_p=3$eV. 
\subsection{Regular octahedron}
We first assume $\alpha=0$, $\beta=90^\circ$ and $D=0$. 
In Table~I(a) we show a set of exchange parameters in the global Hamiltonian 
obtained by setting $U^+=U^-=5$eV and $\Delta=4$ eV. 
Here, we fully consider the energy level splitting of $f^{12}$ in Table \ref{tab1}(a). 
The exchange coupling is very close to the Heisenberg model, i.e. $X\sim Y\sim Z$, 
and when converted to the global axis, they have small but finite off-diagonal terms 
attributed to $\Gamma\ne 0$, while the anti-symmetric part is absent, $D=0$. 
The result is consistent with the calculations given for the triangular lattice geometry\cite{Rau2018}. 
\par
To examine what kind of symmetries or parameters may affect the symmetry of the exchange interactions, 
we consider two effects; 
one is the effect of splitting of $f^{12}$-levels due to crystal field, SOC, 
and Coulomb interactions between two holes, represented by $E_a$ in Table \ref{tab1}. 
If we set $E_a=0$, all 91 levels become degenerate, which is called charging approximation\cite{Onoda2011,Rau2015}. 
The second effect is the influence of lattice symmetry on the hopping term, $\hat T_{\gamma\lambda}$; 
because of the anisotropy of the shape of $f$-orbitals, 
lowering the symmetry will generate inequivalence between the Slater-Koster overlap through 
direction cosines, as discussed in Sec.\ref{sec:slater}. 
In the present framework, the highest symmetry is attained when 
the adjacent octahera are parallel $\alpha=0$ and the S-Yb-S bond angle is $\beta=90^\circ$. 
Table~I(b) shows how these two effects affect the symmetry of $\hat J_{0;b}$. 
When charging approximation holds and both angles keep the highest symmetry, a Heisenberg model is realized. 
While breaking one of them, an XXZ model is realized, and when 
both break down, we find an XYZ model. 
\par
Let us first focus on the first effect. 
To clarify which of the processes contribute to the anisotropy of exchange interactions, 
we plot in Fig.~\ref{f4}(a) separately the contributions from different processes 
to $\Gamma$ normalized by $J^{\rm av}_{b}$ about the data shown in Table~I(a); 
the contributions from processes 2 and 4 (which do not include $f^{12}$ as an excited state), 
and those from processes 1 and 3 separately for 13 different levels of $f^{12}$-state. 
For both 1 and 2 bonds, $\Gamma$ is small, 
whereas the ones from processes 1 and 4 that pass through the four lowest levels have large contributions with different signs, 
which suppresses the amplitude overall. 
As we showed in Table~I(b), this case correspond to $E_a\ne 0$ and $\alpha=0,\:\beta=90^\circ$, 
where the ones associated with 13 levels with finite total angular momentum $J\ne 0$ contribute to the XXZ anisotropy, 
while the two levels having $J=0$ do not. 
To be more precise, $J=6,5,1$ yield large negative $\Gamma$, and $J=4$ have large positive $\Gamma$. 
For large $J$, the two holes polarize their angular momentum $m$. 
The spatial distribution of orbitals is classified into three groups (see Appendix \ref{sec:trans}); 
$m=\pm 3$ and $m=\pm 1$ consists of 
$\cket{x_0(5x_0^2-3r^2)},\,\cket{y_0(5y_0^2-3r^2)}, \cket{x_0(y_0^2-z_0^2)},\,\cket{y_0(z_0^2-x_0^2)}$, 
which have some odd parity about either $x_0$ or $y_0$ within the $x_0y_0$-plane 
(or $C_2$ rotation about the $z_0$-axis).   
On the other hand, $m=\pm 2$ consists of $\cket{x_0y_0z_0}$ and $\cket{z_0(x_0^2-y_0^2)}$, 
which have $C_4$ symmetry about the $z_0$ axis in the $x_0y_0$ plane. 
Otherwise, $m=0$ has no node on the $x_0y_0$-plane, since $\cket{z_0(5x_0^2-3r^2)}$. 
Different $J$ has different constituents between these three groups, 
and their sign influences the sign of hopping integrals, $\hat T_{\gamma\lambda}$. 
The level splitting gives different denominators in the perturbation terms 
and generates inequivalencies between contributions from orbitals of different groups. 
The charging approximation, $E_a=0$ will wipe out this inequivalence and recover the Heisenberg symmetry. 
\par
The strategy to enhance the anisotropy is to decrease $U^-$, which effectively increases the difference 
between $U^-+E_a$ of different levels. 
Figure~\ref{f4}(b) shows the $U^+$ and $U^-$ dependences of $\Gamma$ normalized by $J^{\rm av}_{b}$. 
As we showed in Fig.~\ref{f2}(b), the excited level includes $f^{12}$ and $f^{13}$ simultaneously, and 
has a denominator, $(U^++U^-)$. Therefore the decrease of both $U^\pm$ nearly equally contributes to the enhancement of $\Gamma$. 
When $\Gamma<0$ and large, there appears a strong Ising anisotropy of the Hamiltonian, 
Namely, the pseudo-up/down spins of $\Gamma_6$ doublet 
favor not to change its orientation, because the lowest $J=6$ may dominate. 

\subsection{Tilted octahedron}
The second effect regards the S-Yb-S angle $\beta$, and tilting angle $\alpha$. 
Figure~\ref{f5}(a) shows $\alpha$-dependence of $\Gamma$ and $D/\sqrt{2}$ for bonds 1 normalized by $J^{\rm av}_1$ 
when $U^\pm=5$eV and $\Delta=4$eV. 
Because the inversion symmetry is broken when $\alpha\ne 0$, an antisymmetric Dzyaloshinskii-Moriya term 
becomes finite, and its amplitude grows almost linearly with $\alpha$. 
The $\Gamma$-term is less sensitive and varies parabolically with $\alpha$. 
We plot in the inset the $\beta$-dependence with different choices of $\alpha$. 
The value of $\gamma$ varies similarly to $\beta$ because both are the effect of direction cosines 
in the Slater-Koster overlap, 
and since $\beta$ has not much to do with the inversion symmetry, $D$ is less sensitive than to $\alpha$. 
\par
To know the overall tendency of $\Gamma$, we also plot in Fig.~\ref{f5}(b) 
the ones for bond-1, with different sets 
$(U^+,U^-,\Delta)=(4,4,4),(2,4,4),(2,4,1),(2,2,4)$ and $(2,2,1)$. 
The functional forms are more or less the same, 
and the amplitude seriously depends on $\Delta$; 
those with different $U^\pm$ have smaller differences than those with different $\Delta$. 
This could be because, all the terms include $\Delta$ in part of the perturbation processes, 
which enhances the difference between different processes through the smaller difference in the denominator. 
Namely, the anisotropy here due to $\alpha$ and $\beta$ comes from the numerator, 
and because it is present in the charging approximation, we may safely confine our discussions 
to the hopping from $\Gamma_6$ to $p$-orbitals. 
The $\Gamma_6$ doublet in Eq.(\ref{eq:single-f13}) have both three terms. 
If the $\sigma=+$ spins hop out of this state and come back again, 
it needs to shift from $(m,\sigma)=(0,+)$ to $(-2,-)$ or $(-3,-)$ when it tries to flip the 
orientation of moment, whereas it can come back without doing so when it does not flip. 
Namely, ones with $m=-1,3$ have $C_2$-symmetry about $z_0$, and contribute in different signs 
to the hopping, while $m=0$ does not. 
Therefore, if there is a slight imbalance in the former process because of the distortion, 
the anisotropy appears. 

\section{Summary}
We have derived the super-exchange Hamiltonian, whose spin-1/2 degrees of freedom represent the $4f^{13}$ Kramers doublets on ytterbium atom forming a zigzag chain, with in mind the recently studied magnetic insulator, YbCuS$_2$. 
The local $4f$ Hamiltonian consists of a crystal field from the octahedral ligands, 
Coulomb interactions on $4f$-orbitals, and the spin-orbit coupling, whose complicate energy levels can generate a substantial quantum anisotropy to the super-exchange coupling. 
The fourth-order perturbation processes between the electrons on two Yb ions are mediated by the two S ions on a bond shared by the octahedra, 
and generates a nearly Heisenberg exchange interactions and 
small off-diagonal but symmetric $\Gamma$-interactions. 
The types of anisotropies are systematically understood from the symmetry of the bond Hamiltonian 
in the frame of local $x_0y_0z_0$-axis shared by the two adjacent tetrahedra, which have only diagonal 
interaction but of various symmetries, classified as XXX, XXZ, XYZ. 
These terms are converted to the global $xyz$-coordinate where the diagonal anisotropy transforms to the $\Gamma$-term.  
By analyzing the effect of various model parameters, and by classifying the contributions from different 
perturbation processes, we figured out two important origins of the $\Gamma$ term. 
One is the level splitting of the excited two-hole states of the Yb-ion, 
where a strong Coulomb repulsion between holes and SOC plays a crucial role; 
different levels have different total angular momentum $J$, and the difference in the spatial distribution 
of the corresponding wave functions contribute differently. 
If the effect of level splitting is enhanced, $\Gamma$ increases. 
The other effect is the shape of $\Gamma_6$ doublet, and when the S-Yb-S angle is distorted off 90$^\circ$ 
or the adjacent tetrahedra tilt from parallel configuration, 
the Slater-Koster overlaps of the constituents of these doublets have inequivalence. 
Both the variation of $\alpha$ and $\beta$ generate an anisotropy of $\Gamma$, 
and $\alpha$ that breaks of inversion symmetry yield finite Dzyaloshinskii-Moriya interaction. 
Although many of the previous articles have derived similar effective spin Hamiltonian based on 
perturbation theory, these effects are scarcely highlighted explicitly and systematically. 
\par
The $\Gamma$-term, although its magnitude is smaller by one or two orders of magnitude than the 
Heisenberg term may have a certain influence on the ground state and low energy properties. 
Indeed, without the $\Gamma$ term, the lowest energy excitation of the Heisenberg zigzag chain is a triplet. 
Because $\Gamma$-term supports a finite total spin but zero magnetization, 
it may explain possibly gapless excitation found in the NMR study.

\appendix
\section{$f^{12}$-energy eigenstates of a single ion}\label{sec:f12_2}
In Table \ref{tab1}(a) we show the list of energy eigenstates of ${\cal H}_{\rm ion}$ in Eq.(\ref{eq:hion}) with Eqs.(\ref{eq:hsoc}) and (\ref{eq:hcoulomb}), previosly given in Ref~.\cite{Rau2018}, Table IV. 
There are 91 states and 13 different levels, which are classified by the total angular momentum, $J=l+s$, 
where the degeneracy of each level is given as $2J+1$. 
In Table~\ref{tab1}(b) we show another set of energy levels obtained by ${\mathcal H}_{\rm Coulomb}\rightarrow {\mathcal H}_{\rm Coulomb}/2$, 
where some part of the levels change their orders. 
We performed the same perturbation using (b) and compared it with that by (a), 
and found that a smaller Coulomb interaction between two holes reduces the anisotropy $\Gamma$. 
We analyzed the contributions from different levels as in Fig.~\ref{f4}(a), which are slightly different but have the same tendency. 
However, the level splitting between the lowest two is larger for (a) which gives larger $\Gamma$, 
which is consistent with our observations given in the main text. 
\begin{table}[tb]
\label{tab1}
\caption{Energy levels $E_a$ of $f^{12}$ states of Yb$^{4+}$-ion measured from the lowest, 
obtained by diagonalizing Eq.(\ref{eq:hion}) where we confine the basis to $l=3$. 
We show the result of 
(a) $F_2 = 14.184$eV, $F_4 = 9.846$ eV and $F_6 = 6.890$ eV, $\zeta=0.380$eV (the same as Ref.\cite{Rau2018}) 
and (b) the case where we set $F^a$ to half the value. 
}
\begin{tabular}{clllll}
\hline
(a)&  $E_a$ (eV) & $J$ &(b)& $E_a$ (eV) & $J$ \\
\hline
1& 0.000 & 6 &\;\;& 0.000 & 6 \\
2& 0.752 & 4 && 0.339 & 4 \\
3& 1.187 & 5 && 1.215 & 5 \\
4& 1.785 & 4 && 1.237 & 2 \\
5& 2.040 & 3 && 1.561 & 4 \\
6& 2.094 & 2 && 1.642 & 3 \\
7& 2.985 & 4 && 2.561 & 2 \\
8& 3.924  & 2 && 2.726 & 4 \\
9& 4.751  & 6 && 2.788 & 0 \\
10& 5.023 & 0 && 2.954 & 6 \\
11& 5.167  & 1 && 3.205 & 1 \\
12& 5.396  & 2 && 3.775 & 2 \\
13& 10.558  & 0 && 6.247 & 0 \\
\hline
\end{tabular}
\end{table}
\section{Distortion angles and direction cosines}\label{sec:angles}
We present the relationships between the distortion angles $\alpha$ and $\beta$ and direction cosines 
for the calculation in \S.~\ref{sec:slater}. 
The direction cosines are given for 
Yb-1 $\rightarrow$ S-A , S-A $\rightarrow$ Yb-2, Yb-1 $\rightarrow$ S-B, and 
S-B $\rightarrow$ Yb-2 are indexed as 1,2,3, and 4 and are given as, 
\begin{align}
& (\ell_1,m_1,n_1)= C \bigg(1,-\frac{\tan(\beta/2)}{\cos\alpha},-\tan\alpha\bigg) ,  \nonumber\\
& (\ell_2,m_2,n_2)= C \bigg(1,\frac{\tan(\beta/2)}{\cos\alpha},\tan\alpha\bigg) ,  \nonumber\\
& (\ell_3,m_1,n_3)= C \bigg(1,\frac{\tan(\beta/2)}{\cos\alpha},-\tan\alpha\bigg) ,  \nonumber\\
& (\ell_4,m_1,n_4)= C \bigg(1,-\frac{\tan(\beta/2)}{\cos\alpha},\tan\alpha\bigg), 
\end{align}
with $C=\sqrt{1+(\tan(\beta/2)/\cos\alpha)^2+\tan\alpha^2}$.

\section{Transformation of $f$-orbitals}\label{sec:trans}
The basis we used is indexed by angular and spin momentum as $\mu=(m,\sigma)$ 
while the original degenerate basis 
without considering the crystal field splitting is used in the Slater-Koster tables. 
Here, we only focus on the orbital part, and 
denote a set of annihilation operator as $\bm f^T=(f_{-3,\sigma},f_{-2,\sigma},\cdots, f_{3,\sigma})^T$ 
and the latter annihilation operator for the degenerate basis as  
$\big\{\;\cket{x_0y_0z_0},\,\cket{x_0(5x_0^2-3r^2)},\,\cket{y_0(5y_0^2-3r^2)},\,\cket{z_0(5x_0^2-3r^2)},
\,\cket{x_0(y_0^2-z_0^2)},\,\cket{y_0(z_0^2-x_0^2)},\,\cket{z_0(x_0^2-y_0^2)} \;\big\}$, 
as $\bm g= \{ g_A \}_{A=1,\cdots,7}$. 
The relationships between these two bases are given as $\bm g^T= M \bm f^T$ where 
\begin{align}
M=\left(\begin{array}{ccccccc}
0 & -\frac{i}{\sqrt{2}} & 0 & 0 & 0 & \frac{i}{\sqrt{2}} & 0 \\
\frac{\sqrt{5}}{4} & 0 & -\frac{\sqrt{3}}{4} & 0 &\frac{\sqrt{3}}{4} & 0 & -\frac{\sqrt{5}}{4}\\
\frac{\sqrt{5}i}{4} & 0 & \frac{\sqrt{3}i}{4} & 0 &\frac{\sqrt{3}i}{4} & 0 & \frac{\sqrt{5}i}{4}\\
0 & 0 & 0 & 1 & 0 & 0 & 0 \\
-\frac{\sqrt{3}}{4} & 0 & -\frac{\sqrt{5}}{4} & 0 &\frac{\sqrt{5}}{4} & 0 & \frac{\sqrt{3}}{4}\\
\frac{\sqrt{3}i}{4} & 0 & -\frac{\sqrt{5}i}{4} & 0 &-\frac{\sqrt{5}i}{4} & 0 & \frac{\sqrt{3}i}{4}\\
0 & \frac{1}{\sqrt{2}} & 0 & 0 & 0 & \frac{1}{\sqrt{2}} & 0 
\end{array}\right). 
\end{align}
We find that $m=\pm 3,\pm 1$ are connected to $A=2,3,5,6$ while 
$m=\pm 2$ to $A=1,7$ and $m=0$ is equal to $A=4$. 
These relationships reflect the symmetry of the spatial distributions of orbitals. 

\begin{acknowledgment}
We thank Takahiro Onimaru, Chikako Moriyoshi, Kenji Ishida, Shunsaku Kitagawa, and Fumiya Hori for discussions. 
This work is supported by Grant-in-Aid for Scientific Research (Nos. 21K03440, 21H05191, JP23KJ0783) from the Ministry of Education, Culture, Sports, Science and Technology of Japan. 

\end{acknowledgment}
\bibliographystyle{jpsj}
\bibliography{ybcus2_ref}
\end{document}